\documentclass[12pt,a4paper]{article}
\usepackage[utf8]{inputenc}
\usepackage[plainpages=false,pdfpagelabels=true]{hyperref}
\usepackage{amssymb,amsthm}
\usepackage[margin=1.5 in]{geometry}
\usepackage{graphicx}
\usepackage{a4wide}  
\usepackage{amsfonts}
\usepackage{amsmath}
\usepackage{bbm}
\usepackage{booktabs} 
\usepackage{multicol}

\usepackage{ifpdf}
\ifpdf

\hypersetup{%
  pdftitle    = {The Jacobi metric approach for dynamical wormholes}
  pdfkeywords = {Wormholes, Jacobi metric, geodesics},
  pdfauthor   = {Oscar},
  plainpages  = true,
  colorlinks  = true,
  citecolor   = blue,
  urlcolor    = blue,
  linkcolor   = black
}

\newcommand{\arxiv}[1]{{\tt
\href{http://www.arXiv.org/abs/#1}{#1}}}
\newcommand{\doi}[1]{{\tt DOI:\href{http://dx.doi.org/#1}{#1}}}

\makeatletter
\@addtoreset{equation}{section}
\makeatother

\begin{document}

\begin{center}

{\Large {\bf The Jacobi metric approach for dynamical wormholes}}

\vspace{1.5cm}

\setcounter{footnote}{0}

\renewcommand{\thefootnote}{\alph{footnote}}
{\sl\large  \'{A}lvaro Duenas-Vidal\footnote{E-mail: {alvaro.duenas [at] epn.edu.ec}} and Oscar Lasso Andino}\footnote{E-mail: {oscar.lasso [at] udla.edu.ec}}

\setcounter{footnote}{0}

\vspace{1.5cm}

\vspace{0.3cm}
{\it $^a$ Departamento de F\'isica, Escuela Polit\'ecnica Nacional,\\
Ladr\'on de Guevara E11-253, C.P. 170525, Quito, Ecuador}\\ \vspace{0.3cm}

{\it $^{b}$ Escuela de Ciencias Físicas y Matemáticas, Universidad de Las Américas,\\
Redondel del ciclista, Antigua v\'{i}a a Nay\'{o}n, C.P. 170504, Quito, Ecuador}\\ \vspace{0.3cm}

\vspace{1.8cm}


{\bf Abstract}

\end{center}

\begin{quotation}
We present the Jacobi metric formalism for dynamical wormholes. We show that in isotropic dynamical spacetimes , a first integral of the geodesic equations can be found using the Jacobi metric, and without  any use of geodesic equation. This enables us to reduce the geodesic motion in dynamical wormholes to a dynamics defined in a Riemannian manifold. Then, making use of the Jacobi formalism, we study the circular stable orbits in the Jacobi metric framework for the dynamical wormhole background. Finally, we also show that the Gaussian curvature of the family of Jacobi metrics is directly related, as in the static case, to the flare-out condition of the dynamical wormhole, giving a way to characterize a wormhole spacetime  by the sign of the Gaussian curvature of its Jacobi metric only.
\end{quotation}

\newpage
\pagestyle{plain}


\newpage


\section{Introduction}

The dynamics of massive particles in different spacetimes is of special interest when studying the motion of celestial bodies. By knowing the trajectories of these particles in different geometries we can study some characteristics of the spacetimes where the particle motion is taking place. Moreover, the dynamics of two or more bodies is extremely difficult to solve in a spacetime which is a solution of the Einstein equations. In this sense, any technique helping to solve the equations of motion in a given spacetime is helpful. 

Several methods have been proposed for ``geometrizing'' the dynamics of Hamiltonian systems in classical mechanics. As one of these mechanisms, the Jacobi metric approach helps to  build a Riemannian manifold whose geodesics correspond to the orbits of the original system. The Jacobi approach arises from the Maupertuis principle and, therefore, it works by projecting the dynamics of the system over a previously fixed energy surface. This, of course, restricts \emph{ab initio} the Jacobi approach to autonomous and natural systems.

Applications of the Jacobi approach to gravitational dynamics have been observed from several years ago. In \cite{Pin:1975} the author uses the properties of the Riemannian geometry for solving the dynamics in Newtonian gravity. Using the sign of the Gaussian curvature of the Jacobi metric he was able to classify the trajectories in the two-body problem. Similar results have been found for static spacetimes inside the framework of General Relativity. The Jacobi metric for Schwarzschild spacetime  was first presented in \cite{Gibbons:2015qja}, while the Reissner-Nordström case was solved in \cite{Das:2016opi}. Finally, the Morris-Thorne wormhole was studied using the Jacobi approach in \cite{Arganaraz:2019fup}. Regarding stationary spacetimes, although at first there were some difficulties when applying the Jacobi metric formalism, the problem was solved in \cite{Arganaraz:2021fwu}.  The Jacobi metric approach for static spacetimes has been studied extensively in \cite{Chanda:2016sjg,Chanda:2016aph}. One of the principal results of all these previous works is the fact that it is possible to connect the Lorentzian geometry of  static spacetimes with lower dimensional Riemannian spaces that store information about the timelike geodesic of the firsts and it seems to keep relevant information about spacetime physics \cite{Bera:2019oxg,Arganaraz:2021fwu}.

In spite of the exhaustive study of the spacetime geodesics using the Jacobi metric approach, there are problems that still remain unsolved. In opposition to stationary spacetimes, evolving or dynamical spacetimes are not equipped with a timelike Killing vector. Whereas geodesic dynamics in stationary spacetimes, as we have mentioned,  seems to be well understood in the context of the the Jacobi approach, geodesics in dynamical spacetimes needs to be considered in detail. That is because of geodesic motion in dynamical spacetimes responds to a non-autonomous Hamiltonian and, thus, the energy is not present as a conserved quantity. his lack of the energy as a conserved quantity casts some doubts about the formulation of the Jacobi metric in evolving spacetimes. However, naively we can force the application of the Jacobi approach to any non-autonomous system. This results in a family of Riemannian metrics parametrized by the coordinate time $t$. 

In this article we study the dynamical wormholes\footnote{Sometimes they are called cosmological wormholes.}  \cite{Cataldo:2008ku,Hayward:1998pp} using the Jacobi metric approach. These kind of  wormholes, which are "unified" with black holes \cite{Hayward:1998pp}, have several properties worth of studying\cite{Rehman:2020myc}. Some of the geodesics equations have been solved in \cite{Mishra:2017yrh}, and we must be careful to show that in these particular cases our formalism recovers the already known results.\\
We discovered that the first integral of motion for the null and timelike geodesics can be deduced from the Jacobi metric only, providing a way to determine an expression for $\dot{t}$. Then, we show that the sign of the Gaussian curvature of the Jacobi metric is fixed by the flare-out condition of the wormhole. Therefore, the characterization of the wormhole, namely the existence of a throat, can be inferred directly from the Jacobi metric. More specifically, we obtain a family of Jacobi metrics that are parametrized by an affine parameter, from which the geodesic equations can be deduced.

In section \ref{sec:1} we present the Jacobi metric approach for isotropic dynamical spacetimes and show that, under suitable conditions, there is a conserved quantity which enables us to extend the Jacobi formalism to these spacetimes. Specifically, we show that the extension of the formalism to dynamical spacetimes results in a continuous family of Jacobi metrics. In section \ref{sec:2} we calculate the family of Jacobi metrics for a dynamic wormhole and, from it, we study the circular stable orbits. In section \ref{sec:3} we study the Gaussian curvature of the family of Jacobi metrics and show that its sign is tantamount to the existence of a throat in a spacetime. In section \ref{sec:4} we discuss about the apparent horizon of the metric, showing that they cannot be inherited to the Jacobi metric. Finally, in section \ref{sec:5} we present the discussion.

\section{Jacobi approach for isotropic dynamical spacetimes}\label{sec:1}

We define isotropic dynamical spacetimes as a direct generalization of static ones by adding a scale factor $a(t)$ such that, in a suitable coordinate system $\{t, x^i\}$, the line element is, 
\begin{equation}\label{MetricDynamicalSpaces}
 ds^2 = - e^{2 \Phi(t,x)} dt^2 + a^2(t) h_{ij} dx^i dx^j \equiv g_{\mu\nu} dx^\mu dx^\nu,
\end{equation}
where $h_{ij}(x)$ is a Riemannian metric. By choosing spatial isotropic coordinates the metric \eqref{MetricDynamicalSpaces} becomes
\begin{equation}\label{RiemannianMetric}
 h_{ij} dx^idx^j = f(r) dr^2 +  d\Omega^2_{D-2}, 
\end{equation}
for some differentiable function $f(r)$, where $D$ is the spacetime dimension and $d\Omega^2_{D-2}$ stands  for the  line element over the $(D-2)$-sphere. 

The $t-r$ component of the  associated Einstein field equation is,   
\begin{equation}\nonumber
 R_{tr} = 8 \pi G T_{tr}.
\end{equation}
After some algebra, we get\footnote{We impose the condition $a(t) >0$ in \eqref{MetricDynamicalSpaces} to avoid cosmological singularities.}, 
\begin{equation}\label{EinsteinFieldEqtr}
(D-2) \frac{ a'}{a} \partial_r \Phi = 8\pi G T_{tr},
\end{equation}
where prime means derivative respect to $t$. Then, for solutions without isotropic radiation, i.e. for whose spacetimes with $T_{tr} = 0$, we have two possibilities \cite{Cataldo:2008ku}, 
\begin{enumerate}
 \item $ a' = 0$. This recovers static solutions. 
 \item $\partial_r \Phi = 0$. In this case, we can take $e^{2\Phi(t)} = 1$  by redefining the time coordinate $t$.  
\end{enumerate}
Since we are interested in a general result, where both $a'$ and $\Phi$ are different from zero,  we shall keep the possibility to have isotropic radiation in the spacetime with line element \eqref{MetricDynamicalSpaces}.

A free-falling test particle with mass $m$ in the spacetime \eqref{MetricDynamicalSpaces} responds to the Lagrangian, 
\begin{equation}\label{FreeLagrangian}
 L_{free} = \frac{m}{2} \left[a^2(t) h_{ij} { \dot x}^i { \dot x}^j - {\dot t}^2 e^{2\Phi} \right],
\end{equation}
where the dot means derivative with respect to the proper time $\tau$. We shall call \eqref{FreeLagrangian} the free-falling Lagrangian. The problem with this Lagrangian is the presence of derivatives of the coordinate time $t$. However, we can overcome this difficulty by understanding $t(\tau)$ as an external function to an equivalent one-particle system with Lagrangian \eqref{FreeLagrangian}. Then, defining, 
\begin{equation}\label{definePotandVecpot}
  V = \frac{1}{m} {\dot t}^2e^{2\Phi},
\end{equation}
the free-falling Hamiltonian corresponding to \eqref{FreeLagrangian} takes the form,
\begin{equation}\label{FreeHamil}
 H_{free} = \frac{h^{ij}}{2 m a^2} p_i p_j + m^2 V.  
\end{equation}
This corresponds with the Hamiltonian of an autonomous and  natural equivalent one-particle system, where the potential $V$ depends on $\tau$ through the time function $t(\tau)$. Following this equivalence, over a solution $(x(\tau), p(\tau))$ to the equations of motion we can define the function, 
\begin{equation}\label{FreeEnergy}
 E(\tau) = H_{free}(x(\tau), p(\tau)).
\end{equation}
We shall call $E(\tau)$ the equivalent free-falling energy function.  After some algebra we get
\begin{equation}\label{deltaReatedToFreeEnergy}
 E = \frac{m}{2}\left(\delta + 3 \dot t^2 e^{2\Phi} \right), 
\end{equation}
where $\delta = g_{\mu\nu} \dot x^\mu \dot x^\nu$. Thus $\delta = 0$ for null geodesics and $\delta = \pm 1$ for spatial and timelike geodesics. 

Note that there is no observer related to the measure of $E(\tau)$ and thus, from the point of view of the dynamics in the spacetime \eqref{MetricDynamicalSpaces}, the function \eqref{FreeEnergy} must be viewed as a mathematical artifact without any physical meaning. However, it let us to draw up an equivalent one-particle mechanical system with Hamiltonian \eqref{FreeHamil} and, therefore, from it we can write down the Jacobi metric $J_{ij}$ related to geodesic dynamics in the spacetime \eqref{MetricDynamicalSpaces}, 
\begin{equation}\label{MetricJacobi1}
 J_{ij} = (\delta + \dot t^2 e^{2\Phi}) a(t) h_{ij}. 
\end{equation}
So we do not get a unique Jacobi metric, but a family of continuously parametrized Jacobi metrics by means of $t$. Whereas some mathematical details must be observed with care, the approach is systematically constructed without any obstacle, so we can take \eqref{MetricJacobi1} as a valid result and force the Jacobi approach to dynamical spacetimes.

Since the equivalent free-falling energy is not a bona fide physical energy, but just a useful mathematical function that let us to define an equivalent one-particle classical mechanic system, it is convenient to search for an energy function with physical meaning, i.e. related to an observer. Imposing asymptotic flatness over \eqref{MetricDynamicalSpaces}, $t$ is the time for any asymptotic observer. On the other hand, taking $t$ as the parameter for describing curves in our spacetime, the Lagrangian for time-like geodesics is given by, 
\begin{equation}\label{DefGeoLagran}
\frac{dt}{d\tau} = \frac{1}{L_{geo}},
\end{equation}
After some algebra we get,  
\begin{equation}\label{GeoLagrangian}
 L_{geo}  = [e^{2\Phi} - a^2(t) h_{ij} {x'}^i  {x'}^j]^{1/2}. 
\end{equation}
We shall call \eqref{GeoLagrangian} the geometrical Lagrangian. Defining the momenta $\bar p_i = \frac{\partial L_{geo}}{\partial {x'}^i}$, a Legendre transformation gives us the geometrical Hamiltonian, 
\begin{equation}\label{geometricHamiltonian}
 H_{geo} = -e^{\Phi} \left(1 + \frac{h^{jk}}{a^2(t)} \bar p_j \bar p_k\right)^{1/2}.
\end{equation}
Then, the function, 
\begin{equation}\label{AsymptoticEnergy}
 \bar E(t) = H_{geo}(x(t), \bar p(t)),
\end{equation}
is  the energy measured by an asymptotic observer. We will call \eqref{AsymptoticEnergy} the asymptotic energy function of the geodesic. Note that $\bar E(t)$ is not conserved for dynamical spacetimes because of $H_{geo}$ depends explicitly on $t$. We recover the conservation of $\bar E$ for static spacetimes, when $\dot a = 0$. From \eqref{GeoLagrangian} and \eqref{FreeLagrangian}, $p_i = - m \bar p_i $. Then, from \eqref{AsymptoticEnergy} and \eqref{FreeEnergy},  
\begin{equation}\label{relatedEnergies}
e^{-2\Phi}\bar E^2  - m^2 = m^2\left(\delta + \dot t^2 e^{2\Phi}.  \right) 
\end{equation}
This equation relates the asymptotic energy function $\bar E$ of time-like geodesics with the $\dot t$ function. Therefore, to know the asymptotic energy function $\bar E(t)$ of a geodesic is equivalent to know the differential equation for the $t(\tau)$ function. 

The only obstacle for solving the geodesic motion from \eqref{MetricJacobi1} is the equation for  $\dot{t}$. However, defining the vector field, 
\begin{equation}\label{Dfield}
 \vec{D} = a(t) \partial_t
\end{equation}
after some algebra it is easy to find that,   
\begin{equation} \label{conservedD:1}
 \dot x^\mu \nabla_\mu \left[( \dot x^\nu D_\nu)^2 - \delta\  D^\nu D_\nu\right] = 0, 
\end{equation}
for null geodesics, and for spatial- and timelike geodesics, the last ones whenever $\Phi = 0$. Then, in these cases, 
\begin{equation}\label{conservedD:2}
 K^2 = ( \dot x^\nu D_\nu)^2 - \delta\  D^\nu D_\nu,
\end{equation}
is conserved along geodesics. This provides an equation for $\dot t$ since $ \dot x^\nu D_\nu = a \dot t$. Note that, although $\vec{D}$ coincides with the timelike Killing field for the static case, $ a' = 0$, it is not a Killing field for the more general dynamic case. Indeed, for the static case, $K^2$ coincides with the asymptotic energy of the geodesic.

\bigskip


\section{Geodesic motion in dynamical wormholes from the Jacobi approach}\label{sec:2}
We consider a dynamic generalization of the Morris-Thorne wormhole given by the metric
\begin{equation}\label{worm:1}
ds^2=-e^{2\Phi(r)}dt^2+a^2(t)\left[\frac{dr^2}{\left(1-\frac{b(r)}{r}-\kappa r^2\right)}+r^2d\Omega^2\right],
\end{equation}
 where $d\Omega^2=r^2d\theta^2+r^2\sin^2(\theta)d\phi^2$ and $\kappa=0,1,-1$. From \eqref{MetricJacobi1}, the Jacobi metric for \eqref{worm:1}   is given by,
\begin{equation}\label{jacobin:1}
ds^2=\left(\delta+e^{2\Phi(r)}\dot{t}^2\right)a^2(t)\left(\frac{dr^2}{1-\frac{b(r)}{r}-\kappa r^2}+r^2d\theta^2+r^2\sin^2(\theta)d\phi^2\right).
\end{equation}
Then, dividing by $ds^2$, 
\begin{equation}
1=\left(\delta+e^{2\Phi}\dot{t}^2\right)a^2(t)\left[\frac{1}{1-\frac{b(r)}{r}-\kappa r}\left(\frac{dr}{ds}\right)^2+r^2\left(\frac{d\theta}{ds}\right)^2+r^2\sin^2(\theta)\left(\frac{d\phi}{ds}\right)^2\right].
\end{equation}
Because of spherical symmetry we can work in the equatorial plane $\theta=\frac{\pi}{2}$. Thus, the geodesics of the Jacobi family \eqref{jacobin:1} satisfy, 
\begin{equation}\label{jacobin:2}
1=\left(\delta+e^{2\Phi}\dot{t}^2\right)a^2(t)\left[\frac{1}{1-\frac{b(r)}{r}-\kappa r^2}\left(\frac{dr}{ds}\right)^2+r^2\left(\frac{d\phi}{ds}\right)^2\right].
\end{equation}

To solve \eqref{jacobin:2} we can use the angular momentum conservation law. The Clairaut constant is given by
\begin{equation}\label{clairaut}
l=\left(\delta+e^{2\Phi}\dot{t}^2\right)a(t)^2r^2\left(\frac{d\phi}{ds}\right).
\end{equation}
Replacing \eqref{clairaut} in \eqref{jacobin:2} and after some simplifications we arrive to,
\begin{equation}\label{eqmo:1}
\left(\delta+e^{2\Phi}\dot{t}^2\right)^2\left(\frac{dr}{ds}\right)^2=\left(1-\frac{b(r)}{r}-\kappa r\right)\frac{1}{a(t)^2}\left(\delta+e^{2\Phi}\dot{t}^2-\frac{l^2}{a^2r^2}\right).
\end{equation}
Since the momentum $L = r^2 a^2 \dot{\phi}$ has to be equal to the Clairaut constant $l$, the only way is that,
\begin{equation}
d\tau=\frac{ds}{\delta+e^{2\Phi}\dot{t}^2}.
\end{equation}
That fixes a parametrization for the Jacobi geodesics, and the equation \eqref{eqmo:1} reduces to, 
\begin{equation}\label{eqmo:1.2}
\dot r^2 = \left(1-\frac{b(r)}{r}-\kappa r\right) \frac{1}{a^4}\left(\delta + e^{2\Phi} \dot t^2 - \frac{L^2}{a^2 r^2}\right). 
\end{equation}

From this point, null and timelike geodesics need to be considered separately. For null geodesics, from \eqref{conservedD:2} we get
\begin{equation}\label{dott}
\dot{t}=\frac{K}{a}e^{-2\Phi},
\end{equation}
and the equation for null geodesics reduces to
\begin{equation}\label{eqmo:2}
 \dot r^2 = \left(1-\frac{b(r)}{r}-\kappa r\right) \frac{1}{a^4} \left(K^2 e^{-2 \Phi} - \frac{L^2}{r^2}\right),
\end{equation}
which coincides with the result in \cite{Mishra:2017yrh}. From \eqref{AsymptoticEnergy}, we obtain an asymptotic energy function, 
\begin{equation}\label{energyNull}
 \bar E(t) = \frac{K}{a(t)}. 
\end{equation}
With respect to timelike geodesics, a general ansatz for $\dot t$ can not be provided. However, for lapse function $\Phi = 0$,  from \eqref{conservedD:2} we get
\begin{equation}\label{dott2}
\dot{t}^2=1+\frac{ K^2}{a^2}.
\end{equation}
Then, the geodesic equation reduces to, 
\begin{equation}\label{eqmo:2timelike}
 \dot r^2 = \left(1-\frac{b(r)}{r}-\kappa r\right) \frac{1}{a^4} \left(K^2 - \frac{L^2}{r^2}\right).
\end{equation}
which also coincides with the result in \cite{Mishra:2017yrh}, and we obtain for the asymptotic energy function, 
\begin{equation}\label{energyTimelike}
 \bar E^2(t) = m^2 + \frac{K^2}{a^2(t)}. 
\end{equation}
Therefore, we have recovered the equations for null and timelike geodesics in the spacetime \eqref{worm:1} without the use of geodesic equation for \eqref{worm:1}. Thus, as we expected, the whole information regarding the geodesics of \eqref{worm:1} is stored in \eqref{jacobin:1}.

\subsection{Circular stable orbits}

Here we study the circular stable orbits coming from \eqref{eqmo:2}. To keep the first integral \eqref{conservedD:2} for timelike geodesics, we set $\Phi=0$. Also, for simplicity, we rewrite equation \eqref{jacobin:1} using the change of variable,
\begin{equation}
\rho=a(t)r,
\end{equation}
then, a general shape function $b(r)=b_{o}^{n}r^{-n}$ transforms to 
\begin{equation}
b(\rho)=b^n_ou^{n-1}a^{n-1}
\end{equation} 
The equation of circular orbits in the $u$ variable takes the form, 
\begin{equation}\label{ufinal2}
\left(\frac{du}{d\phi}\right)^2=\frac{1}{a^2(t)}\left(1-b(\rho)a(t) u-\kappa \frac{1}{u^2a^2(t)}\right)\left(\delta+K^2-Lu^2a^2(t)\right).
\end{equation}
Taking $\delta=0$ we have\footnote{$\delta=0$ for null geodesics, although we know that for timelike geodesics the equation of motion is going to be the same.},
\begin{equation}\label{eq:1}
a^{2}(t)\left(\frac{du}{d\phi}\right)^2=\frac{K^2}{L^2}\left(1-a^n(t)b_o^n u^{n}-\kappa \frac{1}{u^2a^2(t)}\right)\left(1-\frac{L^2}{K^2}a(t)^2u^2\right).
\end{equation}
The previous equation can be written as,
\begin{equation}
\left(\frac{du}{d\phi}\right)^2=f(u),
\end{equation}
where,
\begin{equation}
f(u)=\frac{1}{a^2(t)}\frac{K^2}{L^2}\left(1-a^n(t)b_o^n u^{n}-\kappa \frac{1}{u^2a^2(t)}\right)\left(1-\frac{L^2}{K^2}a^2(t)u^2\right).
\end{equation}
Then, for a circular orbit the function $f(u)$ has to satisfy,
\begin{eqnarray}
f(u)&=&0\label{cir:1}\\
f'(u)&=&\frac{2 k}{a^2 u^3 C} - 2 a^2 u - \frac{a^n bo^n n u^{n-1}}{C} + a^{n+2} bo^n (2 + n) u^{n+1}=0\label{cir:2}
\end{eqnarray}
The system \eqref{cir:1}, \eqref{cir:2} leads to the condition,
\begin{equation}
\kappa+a^2u^2(a^nb_{o}^nu^n-1)=0
\end{equation}
which for $k=0$ reduces to,
\begin{equation}
u_{c}^n=\frac{1}{a^n b_{o}^n}.
\end{equation}\\
The previous condition implies that the only stable circular orbit is located at the throat of the wormhole\footnote{For static wormholes this condition is reduced to $u_{c}^n=\frac{1}{b_{o}^n}$ which implies that the only circular orbit is located at the throat of the wormhole \cite{Arganaraz:2019fup}.}. We want to see if this orbit is stable. In order to do so we can determine (for $\kappa=0$) the sign of $f''(u)$. Thus,
\begin{equation}
f''\left(u=u_c,C=\frac{1}{a^n b_o^n }\right)=-4a^2n<0,
\end{equation}
therefore, the circular orbit located at the throat is stable.

When $\kappa=1$ we only have solutions for the $n=2$ case
\begin{eqnarray}
u_{c}&=&\pm\frac{1}{a(t)}\sqrt{\frac{2}{1+ \sqrt{(1 - 4 bo^2)}}}\\
u_{c}&=&\pm\frac{1}{a(t)b_o}\sqrt{\frac{1+ \sqrt{(1 - 4 bo^2)}}{2}}
\end{eqnarray}\\
For ($\kappa=-1$) we get,
\begin{equation}
u_{c}=\pm\frac{1}{a(t)b_o}\sqrt{\frac{1\pm \sqrt{1 + 4 bo^2}}{2}}\\
\end{equation}
From the previous equations we can conclude that the circular orbits of the metric \eqref{worm:1} are located at the throat, and since the wormhole is evolving with time the length of the throat is changing. The way of how the throat changes is given by the function $a(t)$. Moreover, the location of the throat changes for the cases $\kappa=\pm 1$ and we are able to determine them only when $n=2$.

\subsection{The $\kappa=0$  case for the $n=2$ wormhole}

\begin{figure*}
\begin{multicols}{2}
    \includegraphics[scale=0.8]{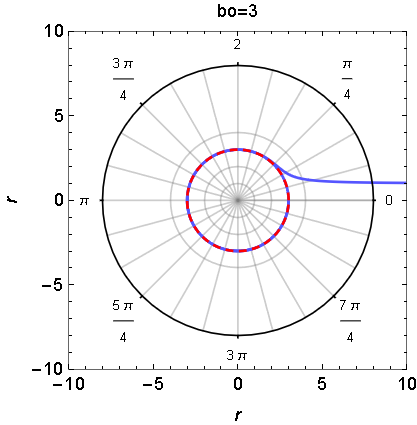}\par 
     
   \includegraphics[scale=0.8]{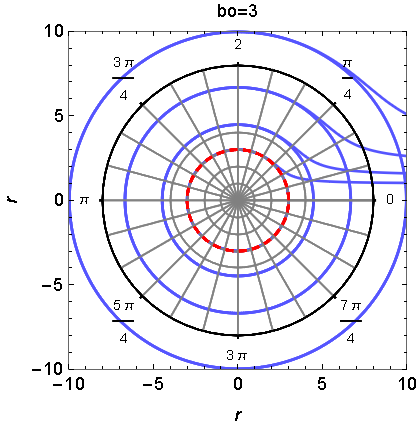}\par 
     
    \end{multicols}
    \caption{Polar plot of a trajectory of a particle coming form infinity, throat size $bo=3$. We have used equation (\ref{equat}). The blue lines represent trajectories with the only condition $C=1/(a(t)^2bo^2)$. This graph shows the only circular stable orbit located at the throat at $t=0, 1/5, 2/5, 3/5, 4/5, 1, 6/5, 7/5$. The red line represents the orbit $C=1/bo^2$ which corresponds to the static Morris-Thorne wormhole.}
    \label{closedyna:1}
\end{figure*}

Here we study the orbits of the $n=2$ dynamic wormhole with $\kappa=0$. Thus, equation \eqref{eq:1} becomes, 
\begin{equation}\label{root}
\left(\frac{du}{d\phi}\right)^2=\frac{1}{a^2(t)}\frac{K^2}{L^2}\left(1-a^2(t)b_o^2 u^{2}\right)\left(1-\frac{L^2}{K^2}a(t)^2 u^2\right)=(u-\alpha)(u-\beta)(u-\gamma)(u-\eta),
\end{equation}
 where $\alpha,\beta,\gamma,\eta$ are the the roots of $f$. After comparison we obtain the following system:
 \begin{eqnarray}
 \alpha\beta\gamma\eta&=&\frac{1}{C},\\
 \alpha\beta\gamma+ \alpha \beta \eta + \alpha \gamma \eta + \beta \gamma \eta&=&0,\\
 \alpha \beta + \alpha \gamma + \beta \gamma + \alpha \
\eta + \beta \eta + \gamma \eta&=&a^2(1-\frac{b_o^2}{C}),
 \end{eqnarray}
 
 which is solved by
 \begin{equation}
 \alpha=1,\,\,\,\beta=-1,\,\,\,\gamma=-\frac{1}{\sqrt{C}},\,\,\,\eta=\frac{1}{\sqrt{C}}.
 \end{equation} 
 Hence, the radial trajectories are solved by,
 \begin{equation}
 \frac{du}{\sqrt{(u-\alpha)(u-\beta)(u-\gamma)(u-\eta)}}=\frac{1}{a^2(t)}d\phi.
 \end{equation}
 The solution to the previous equation is given by, 
 \begin{equation}\label{equat}
r(\phi)=b_{o}a^2(t)\cos\left(sn\left(\sqrt{b_{o}^2\lvert C\rvert+1}\,\,\phi,\frac{b_{o}\sqrt{\lvert C \rvert}}{\sqrt{b_{o}^2\lvert C\rvert+1}}\right) \right)
\end{equation}
 In Fig. \eqref{closedyna:1} we present a polar plot for the function $r(\phi)$. We have chosen\footnote{The form factor $a(t)$ comes from the Friedman equation. We have set all constants equal to one.} $a(t)=\exp(t)$. The behavior of the only stable orbit is similar to the static case, this orbit is located always at the throat. The dynamic wormhole has a throat at $r_o$ where $b(r_o)=ro$. But the overall size of the wormhole is changing and therefore the size of the throat is also changing. In Fig. \eqref{closedyna:1}  the blue lines are the trajectories of a particle coming from infinity and describing  the only circular stable trajectory at the throat.

In figure \eqref{closedyna:1}, in the left panel we have plotted the only circular stable orbit in blue. In the right panel we plot circular stable orbits for $t=0, 1/5, 2/5,3/5,4/5,1,6/5,7/5$. All of them are at the throat.  What we see is a very similar behavior compared with the static case.

\section{The flare-out condition and the Gaussian curvature}\label{sec:3}

In \cite{Pin:1975} the sign of the Gaussian curvature of the Jacobi metric was used to classify the trajectories in the Kepler problem. Here we want to see how this sign is related to the flare-out condition of the wormhole. In \cite{Arganaraz:2019fup} it was shown that indeed the Gaussian curvature has a sign determined by the flare-out condition of the static wormhole. In this section we show that for the dynamic wormhole \eqref{worm:1} we have a similar relation. However, there are some subtleties that we will discuss regarding the function $a(t)$. Thus, the Gaussian curvature of a metric of the form,
\begin{equation}
ds^2=f^{2}\left(\frac{dr^2}{g^2(r)}+r^2d\phi^2\right),
\end{equation}
where,
\begin{equation}
f^2=\delta-E^2a^2(t),\,\,\,\,\,\,\,g(r)^2=1-\frac{b(r)}{r}-\kappa r^2,
\end{equation}
is given by,
\begin{equation}\label{gaussian:1}
\frak K=\frac{r b'(r)-b(r)+2\kappa r^3}{2 r^3\left(E^2-\delta a^{2}(t)\right)}.
\end{equation}
It turns out that the numerator is the flare-out condition of the wormhole \eqref{worm:1} and it has a definite sign. Therefore, if $E^2>\delta a^2(t)$ the Gaussian curvature is negative at the throat. The condition for the energy and the function $a(t)$ deserves more study. As we expected, the Gaussian curvature, which is an intrinsic quantity, is related to the flare-out condition which is a coordinate dependent quantity. We understand that because of the Jacobi metric projection the time coordinate has disappeared and therefore the old coordinate $t$ parametrizes a family of Jacobi metrics. The  expresion in \eqref{gaussian:1} is a family of Gaussian curvatures parametrized by $t$ through the function $a(t)$. When $a(t)=1$ we recover the static case for an asymptotically AdS wormhole. When $a(t)=1$ and $\kappa=0$ we recover the static case studied in \cite{Arganaraz:2021fwu}. Certainly, the existence of a throat is inherited to the Jacobi metric. Although the determinant characteristic of a wormhole is the throat, we will see that there are some physical quantities which are not inherited to the Jacobi metric. These quantities are related to the apparent horizons of the metric \eqref{worm:1}.

\section{About the horizons}\label{sec:4}

As we have studied, the Jacobi metric does inherit the geodesic properties of the original spacetime metric. However, there are physical quantities which are completely lost. In static or stationary spacetimes it was shown in \cite{Arganaraz:2021fwu} that the localization of the horizon was inherited to the $2-$dimensional generalized Jacobi metric.  In this section we study the the apparent and trapped horizons. We argue that the information regarding both type of  horizons disappear when the Jacobi metric was calculated.
A spacetime with spherical symmetry can be foliated by $2-$spheres . If we define the coordinates over the a $2-$ sphere as $(x,y)$ the condition needed for a $2-$sphere to be a trapped surface is 
\begin{equation}\label{exp:1}
\frac{\partial r(x,y)}{\partial x}<0,\,\,\,\,\,\,\,\,\,\,\,\,\,\,\frac{\partial r(x,y)}{\partial y}<0.
\end{equation}
The condition \eqref{exp:1} says that the outgoing and ingoing null geodesics normal to $S$ have negative expansions. The boundary of a trapped region is called a trapped horizon, and a spacelike slicing of the trapping horizon is denominated an apparent horizon.\\
In order to find the position of the trapped surfaces the null coordinates are used. These coordinates are given by
\begin{equation}\label{null:1}
x^{+}=t+r_{*},\,\,\,\,\,\,\,\,\,x^{-}=t-r_{*}.
\end{equation}
The metric \eqref{worm:1} written using the coordinates \eqref{null:1} and taking, 
\begin{equation}
\frac{dr}{dr_{*}}=\frac{1}{a(t)}\sqrt{1-\frac{b(r)}{r}},
\end{equation}
becomes,
\begin{equation}
ds^2=2g_{+-}dx^{+}dx^{-}+R^{2}d\Omega^{2},
\end{equation}
where now $R=R(r^{+},r^{-})=a(t)r$ is the areal radius and $g_{+-}=-1/2$. The element $d\Omega^2$ is the metric for the $2-$dimensional sphere.
Using the definition of expansion as $\Theta_{\pm}=\frac{2}{R}\partial_{\pm}R$ we can determine a criteria for a trapped surface, thus a sphere is trapped if $\Theta_{+}\Theta_{-}>0$ and untrapped if $\Theta_{+}\Theta_{-}<0$. Finally, a sphere is marginal if $\Theta_{+}\Theta_{-}=0$. A trapping horizon is determined by the condition $\Theta_{+}|_{h}=0$. This condition leads, for the trapping horizon $R_{h}=a(t)r_{h}$, to,
\begin{equation}
\dot{R_{h}}+\sqrt{1-\frac{a(t)b(r)}{R_{h}}}=0.
\end{equation}
The trapped horizon is future (past) if $\Theta_{-}<0\,(\Theta_{-}>0)$ and bifurcating if $\Theta_{-}=0$. Thus, when we calculate the Jacobi metric the areal radius remains untouched, but it has lost its physical significance. The metric \eqref{jacobin:1} together with \eqref{dott} define a three dimensional spherically symmetric space metric\footnote{Note that the Jacobi metric is a Riemannian metric} which can be foliated by $2-$dimensional spheres. However, the idea of trapped surface is lost. The definition of a Kodama vector is lost and we cannot study the whole thermodynamics of the wormhole only by using the family of Jacobi metrics. 

\section{Discussion}\label{sec:5}

As we have shown, the Jacobi formalism can be extended to study isotropic dynamical spacetimes. Under some considerations (null geodesics or timelike geodesics for $\Phi = 0$),  a first integral of motion can be found without making use of the geodesic equation. This give us an equation for $\dot t$, and the formalism results in a continuous family of Jacobi metrics. 

The formalism has been checked over the dynamical wormhole \eqref{worm:1}, and we have obtained some results that were expected. Specifically, 
from the Jacobi approach we have shown that the only stable circular orbit is located at the throat. Since the size of the wormhole is changing with the form factor $a(t)$, the size of the throat is changing and therefore the size of the stable circular orbit also does.

In figure \eqref{closedyna:1}, in the left panel we have plotted the only circular stable orbit in blue. In the right panel we plot circular stable orbits for $t=0, 1/5, 2/5,3/5,4/5,1,6/5,7/5$. All of them are at the throat.  What we see is a very similar behavior compared with the static case. We have to remember that in the dynamical case there is not a surface energy were we can project, but using the formalism we have developed here,  we are able to find and study the corresponding family of Jacobi  metrics.

We were able to show that the property of being a wormhole, namely the throat, is inherited to the family of Jacobi metrics. The Gaussian curvature of the family of Jacobi metrics has a definite sign, being related to the flare-out condition  and,  therefore, it shows the presence of the throat. In this way, we would only need to know the family of Jacobi metrics in order to determine if a dynamical spacetime contains a wormhole. A similar result was obtained for static wormholes in \cite{Arganaraz:2019fup} and here we have generalized to the dynamic case. 

Finally, naively we have forced the application of the Jacobi approach to non-autonomous systems, obtaining a family of Riemannian metrics parametrized by the coordinate time $t$. This has worked when we look for a Jacobi metric, however it is not clear how the projection is mathematically defined. This  must be worked out if we want a mathematical formulation of the Jacobi metric approach for these type of systems. Here we have shown that the procedure works, but some information is lost, such as the trapped horizons. We hope that with a mathematical formulation we can clarify this aspect. Thus an extension of the Jacobi metric approach to non-autonomous systems can be proposed as a future work.





\begin{thebibliography}{9}

\bibitem{Pin:1975}  
O.~Chong Pin, "Curvature and mechanics", Advances in Mathematics, 15, 3, 269-311, (1975),

\bibitem{Gibbons:2015qja}
G.~W.~Gibbons,
``The Jacobi-metric for timelike geodesics in static spacetimes,''
Class. Quant. Grav. \textbf{33} (2016) no.2, 025004,
\doi{10.1088/0264-9381/33/2/025004},
arXiv: \arxiv{1508.06755} [gr-qc].

\bibitem{Das:2016opi}
P.~Das, R.~Sk and S.~Ghosh,
``Motion of charged particle in Reissner\textendash{}Nordstr\"om spacetime: a Jacobi-metric approach,''
Eur. Phys. J. C \textbf{77} (2017) no.11, 735,
\doi{10.1140/epjc/s10052-017-5295-6},
arXiv:\arxiv{1609.04577} [gr-qc].

\bibitem{Arganaraz:2019fup}
M.~Arga\~naraz and O.~Lasso Andino,
``Dynamics in wormhole spacetimes: a Jacobi metric approach,''
Class. Quant. Grav. \textbf{38} (2021) no.4, 045004,
\doi{10.1088/1361-6382/abcf86},
arXiv:\arxiv{1906.11779} [gr-qc].

\bibitem{Arganaraz:2021fwu}
M.~A.~Arga\~naraz and O.~L.~Andino,
``The generalized Jacobi metric,''
arXiv:\arxiv{2112.10910}[gr-qc].


\bibitem{Chanda:2016sjg}
  S.~Chanda, G.~W.~Gibbons and P.~Guha,
  ``Jacobi–Maupertuis metric and Kepler equation,''
  Int.\ J.\ Geom.\ Meth.\ Mod.\ Phys.\  {\bf 14} (2017) no.07,  1730002,
 arXiv:\arxiv{1612.07395} [math-ph].

\bibitem{Chanda:2016aph}
  S.~Chanda, G.~W.~Gibbons and P.~Guha,
  ``Jacobi-Maupertuis-Eisenhart metric and geodesic flows,''
  J.\ Math.\ Phys.\  {\bf 58} (2017) no.3,  032503
 arXiv:\arxiv{1612.00375}.
 


\bibitem{Bera:2019oxg}
A.~Bera, S.~Ghosh and B.~R.~Majhi,
``Hawking radiation in a non-covariant frame: the Jacobi metric approach,''
Eur. Phys. J. Plus \textbf{135} (2020) no.8, 670,
\doi{10.1140/epjp/s13360-020-00693-1},
arXiv:\arxiv{1909.12607} [gr-qc].





\bibitem{Hayward:1998pp}
S.~A.~Hayward,
``Dynamic wormholes,''
Int. J. Mod. Phys. D \textbf{8} (1999), 373-382,
\doi{10.1142/S0218271899000286},
arXiv:\arxiv{gr-qc/9805019}[gr-qc].


\bibitem{Rehman:2020myc}
M.~Rehman and K.~Saifullah,
``Thermodynamics of dynamical wormholes,''
JCAP \textbf{06} (2021), 020,
\doi{10.1088/1475-7516/2021/06/020},
arXiv:\arxiv{2001.08457} [gr-qc].


\bibitem{Cataldo:2008ku}
M.~Cataldo, S.~del Campo, P.~Minning and P.~Salgado,
``Evolving Lorentzian wormholes supported by phantom matter and cosmological constant,''
Phys. Rev. D \textbf{79} (2009), 024005,
\doi{10.1103/PhysRevD.79.024005},
arXiv:\arxiv{0812.4436} [gr-qc]].

   

  

\bibitem{Mishra:2017yrh}
A.~Mishra and S.~Chakraborty,
``On the trajectories of null and timelike geodesics in different wormhole geometries,''
Eur. Phys. J. C \textbf{78} (2018) no.5, 374,
\doi{10.1140/epjc/s10052-018-5854-5},
arXiv:\arxiv{1710.06791}[gr-qc].

  




\end{thebibliography}
\end{document}